\documentstyle[prl,aps,preprint]{revtex}

\begin{document}
\tightenlines
\draft
\widetext

\title{Supersymmetric FRW model and the ground state of supergravity}
\author{ V.I. Tkach$^a$ \thanks{E-mail: vladimir@ifug1.ugto.mx},
J.J. Rosales $^{a,b}$\thanks{E-mail: juan@ifug3.ugto.mx}, 
and J. Socorro $^a$\thanks{E-mail: socorro@ifug4.ugto.mx}\\ 
$^a$Instituto de F\'{\i}sica de la Universidad de Guanajuato,\\
Apartado Postal E-143, C.P. 37150, Le\'on, Guanajuato, M\'exico\\
$^b$ Ingenier\'{\i}a en computaci\'on, Universidad del Baj\'{\i}o\\
Av. Universidad s/n Col. Lomas del Sol, Le\'on, Gto., M\'exico}

\maketitle
\date{\today}

\begin{abstract}
In this work we construct the vacuum configuration of supergravity 
interacting with homogeneous complex scalar matter fields. The 
corresponding configuration is of the FRW model invariant under the 
$n=2$ local conformal time supersymmetry, which is a subgroup of the four 
dimensional space-time supersymmetry. We show, that the potential of the 
scalar matter fields is a function of the K\"ahler potential and of the 
arbitrary parameter $\alpha$. This parameter enumerates the vacuum states. 
The scalar matter potential induces the spontaneous breaking of 
supersymmetry in supergravity. On the quantum level our model is a 
specific supersymmetric quantum mechanics, which admits quantum states 
in supergravity, and the states with zero energy are described by the 
wave function of the FRW universe. 

\end{abstract}

\narrowtext
\newpage
\section {INTRODUCTION}

There are several reasons for studying local supersymmetric theories
rather than non-supersymmetric ones. Local gauge supersymmetry or 
supergravity leads to a constraint, which can be thought of as the 
``square root" of the Wheeler-DeWitt constraint, and it is related to it 
in the same way as the Dirac equation is related to the Klein-Gordon 
equation, this fact has appeared in \cite{uno}. Starting with Hamiltonian
treatment of classical supergravity \cite{dos,tres,cuatro} 
the canonical quantum theory of supergravity has appeared in \cite{cinco} 
using two-component spinor notation. To obtain quantum physical 
solutions of the general theory of supergravity is a laborious 
assignment, this is due to the fact, that in the theory there are 
infinite degrees of freedom. One option to the search of quantum physical 
solution is the so-called supersymmetric minisuperspace models 
\cite{seis,siete,ocho,nueve}, which suppose that space-time is 
homogeneous. In other words, the gravitational and matter fields have 
been reduced to a finite number of degrees of freedom. 
Thus, the study of the associated quantum supersymmetric cosmology 
becomes analogous to a supersymmetric quantum mechanical problem. The 
hope we have in these models is that they could give us the notion of the 
full quantum theory of supergravity. However, not all the results 
obtained in quantum supersymmetric cosmology have had their countrapart 
in the full theory of supergravity. Some of these problems have 
already been presented in two comprehensive and 
organized works: a book and an extended review \cite{diez}.

In the last three years we have proposed a new approach to investigate 
the supersymmetric quantum cosmology \cite{once}. The main 
idea is as follows: one knows that the action of the cosmological models 
obtained from four dimensional Einstein-Hilbert action by spatially 
reduction preserves the invariance under the local time reparametrization. 
Hence, the more extended symmetry must be local. In order to have a more 
extended symmetry in \cite{once} the odd ``time" parameter $\eta$ and 
their complex conjugate $\bar\eta$ were introduced, which 
are the superpartners of the usual time parameter $t$. Then, the new 
functions become superfunctions depending on $t, \eta, \bar\eta$, which 
in the field theory are called superfields. This generalization of the 
local time reparametrization is well-known in the new formulation of the 
spinning particles and superparticles \cite{doce,trece}. This 
procedure allowed us to formulate in the superfield representation the 
supersymmetric action for all the Bianchi-type cosmologies \cite{catorce}, 
which are invariant under the $n=2$ local conformal time supersymmetry 
transformations. The inclusion of the real scalar supermatter fields, as 
well as the spontaneous breaking of local supersymmetry, were obtained in 
\cite{quince} following this procedure.  

In the previous work \cite{diesiseis} we have constructed the most general 
interacting action for the supersymmetric FRW model with a set of 
spa\-tia\-lly ho\-mo\-ge\-neous complex scalar matter superfields. 
This scheme of interaction gives a general mechanism of spontaneous 
breaking of the local supersymmetry analogous to the case of interaction 
of supergravity with chiral matter in four space-time dimensions 
\cite{diesisiete}. Following these investigations in the present work 
we have shown, that in the quantum version the supersymmetric action 
in \cite{diesiseis} describes the vacuum configuration of supergravity 
interacting with homogeneous complex scalar matter supermultiplets. The 
corresponding configuration is of the FRW model invariant under the 
$n=2$ local supersymmetry, which is a subgroup of the four space-time 
supersymmetry. 

The steps to follow this line of research are: 1) the construction of the
general-type superfield action interacting with two superfunctions of the 
kinetic term $\Phi(Z^A, \bar Z^{\bar A})$ and of the superpotential 
$g(Z)$; 2) Weyl rescaling of the superfields; 3) recombination of the 
function $\Phi(Z^A, \bar Z^{\bar A})$ and $g(Z)$ in the K\"ahler 
superfunction $G(Z^A, \bar Z^{\bar A})$; 4) elimination of the auxiliary 
fields and analisys of spontaneous breaking of supersymmetry and 
5) classical and quantum Hamiltonian and supercharges.

The first three steps may be realized if we see, that the simplest way to 
construct the classical Lagrangian is considering the superfields on the 
superspace $(t, \eta, \bar\eta)$. This classical Lagrangian describes the 
evolution of the bosonic and additional Grassmann degrees of freedom, which 
after quantization become generators of the Clifford algebra.
 
In the fourth step we will see, that the scalar field potential is 
described by K\"ahler function and by one arbitrary parameter. We show, 
that we have not one parametrical family theory with new parameter, and 
we have a family of vacuum states in supergravity invariant under $n=2$ 
conformal supersymmetry, which is subgroup of space-time supersymmetry.    

The plan of this paper is the following: in the second section 
the $n=2$ local superfield formulation of the FRW model interacting with 
a set of spatially homogeneous complex scalar matter superfields 
\cite{diesiseis} is reviewed in order to fix notation. In the third section 
the supersymmetric lagrangian without auxiliary fields is written, and the 
spontaneous breaking of local supersymmetry is analized. In the section 
four we consider the canonical formalism on the classical level, and 
the fundamental supersymmetric charges and the Hamiltonian are explicitly 
written. Section five is devoted to discussion of quantization in 
particular, we will see that the supersymmetric still allows ambiguity on 
the factor ordering of the quantum operators, as well as we note, that 
the specific supersymmetric quantum mechanics of our model is due to the 
fact, that the particles-like excitation doesn't correspond to the scale
factor $R$. 
     
\section {SUPERFIELD FORMULATION AND SYMMETRY}

We begin with the superfield action \cite{diesiseis} 
\begin{eqnarray}
S &=& \int \left \{\Phi \left [{I\!\!N}^{-1} {I\!\!R} 
D_{\bar \eta} {I\!\!R} D_\eta {I\!\!R}  - \sqrt{k} {I\!\!R}^2 
- \frac{1}{2} \left \{D_{\bar \eta} \left({I\!\!N}^{-1} {I\!\!R}^2 
D_\eta {I\!\!R} \right) - D_\eta \left({I\!\!N}^{-1} {I\!\!R}^2 
D_{\bar \eta} {I\!\!R} \right ) \right \} \right] \right. \nonumber\\ 
&+& \left. \frac{1}{4}{I\!\!N}^{-1} {I\!\!R}^3 \Phi^{-1} D_{\bar \eta} 
\Phi D_{\eta} \Phi + \frac{1}{2\alpha} {I\!\!N}^{-1} {I\!\!R}^3 
\frac{\partial^2 \Phi}{\partial \bar Z^{\bar A} \partial Z^B} 
\left [D_{\bar \eta} \bar Z^{\bar A} D_\eta Z^B + 
D_{\bar \eta} Z^B D_\eta \bar Z^{\bar A} \right ] \label{uno} \right. \\
&-& \left. \frac{1}{2\alpha} {I\!\!N}^{-1} {I\!\!R}^3 
\Phi^{-1} \frac{\partial \Phi}{\partial \bar Z^{\bar A}} 
\frac{\partial \Phi}{\partial Z^B} \left [D_{\bar\eta} \bar Z^{\bar A} 
D_\eta Z^B + D_{\bar \eta} Z^B D_\eta \bar Z^{\bar A} \right ] 
- \frac{2 {I\!\!R}^3}{\kappa^3} |g(Z)|^{\alpha} \right \} 
d \eta d \bar \eta dt, \nonumber
\end{eqnarray}
with $k=0,1$ stands for plane and spherical FRW, and $\kappa^2 = 8\pi G_N$, 
where $G_N$ Newton's constant of gravity and 
$\alpha$ is an arbitrary constant parameter. As it will be shown, this 
parameter is not fixed by the local conformal time supersymmetry. We can 
see from (\ref{uno}), that the interaction depends on two arbitrary 
superfunctions $\Phi(Z^A, \bar Z^{\bar A})$ and $g(Z^A)$, which is the 
dimensionless superpotential. Making the following Weyl conformal 
transformations

\begin{eqnarray}
&&{I\!\!N} \to  \exp (\frac{\alpha {I\!\!K}}{6}) {I\!\!N}\, , \qquad
{I\!\!R} \to  \exp (\frac{\alpha {I\!\!K}}{6}) {I\!\!R}, 
\label{weyl}\\
&&\Phi \exp (\frac{\alpha {I\!\!K}}{3}) = -\frac{3}{\kappa^2}, \nonumber
\end{eqnarray}
we find, that the terms standing under the integration of (\ref{uno})
take the form  

\begin{eqnarray}
&&\Phi \left [{I\!\!N}^{-1}{I\!\!R} D_{\bar \eta}{I\!\!R} 
D_{\eta}{I\!\!R} - \sqrt k {I\!\!R}^2 - \frac{1}{2}\lbrack 
D_{\bar \eta}({I\!\!N}^{-1}{I\!\!R}^2
D_{\eta}{I\!\!R}) - D_{\eta}({I\!\!N}^{-1}{I\!\!R}^2 
D_{\bar \eta}{I\!\!R}) \rbrack \right ]  \nonumber \\ 
&=& - \frac{3}{\kappa^2}{I\!\!N}^{-1}{I\!\!R} D_{\bar \eta}{I\!\!R}
D_{\eta}{I\!\!R} + \frac {3\sqrt k}{\kappa^2}{I\!\!R}^2 + 
\frac{\alpha^2}{12\kappa^2} {I\!\!N}^{-1} {I\!\!R}^3 D_{\bar \eta}{I\!\!K}
D_{\eta}{I\!\!K} \label{dos} \\
&&\mbox{}- \frac{1}{2} D_{\bar \eta} \left [\frac{\alpha}{4\kappa^2}
{I\!\!N}^{-1} {I\!\!R}^3 D_{\eta} {I\!\!K} + \frac{3}{2\kappa^2} 
{I\!\!N}^{-1} {I\!\!R}^2 D_{\eta} {I\!\!R} \right ] \nonumber\\
&& \mbox{} - 
D_{\eta} \left [\frac{\alpha}{4\kappa^2} {I\!\!N}^{-1} {I\!\!R}^3 
D_{\bar \eta} {I\!\!K} + \frac{3}{2\kappa^2} {I\!\!N}^{-1} {I\!\!R}^2 
D_{\bar \eta} {I\!\!R} \right ],  \nonumber
\end{eqnarray} 

\begin{eqnarray}
\frac {1}{4} {I\!\!N}^{-1}{I\!\!R}^3 \Phi^{-1} D_{\bar \eta} \Phi D_{\eta} 
\Phi 
&=& -\frac{\alpha^2}{12\kappa^2}{I\!\!N}^{-1}{I\!\!R}^3 
D_{\bar \eta}{I\!\!K}D_{\eta}{I\!\!K}, \label{tres}
\end{eqnarray}

\begin{eqnarray}
&& \frac {1}{2\alpha} {I\!\!N}^{-1} {I\!\!R}^3 
(\frac{\partial^2 \Phi} {\partial \bar Z^{\bar A} \partial Z^B} - 
\Phi^{-1} \frac {\partial \Phi}{\partial \bar Z^{\bar A}} 
\frac{\partial \Phi}{\partial Z^B}) \left [ D_{\bar \eta} 
\bar Z^{\bar A} D_{\eta} Z^B + D_{\bar \eta}Z^B D_{\eta} \bar Z^{\bar A} 
\right ] \label{cuatro} \\
&=& \frac {1}{2\kappa^2}{I\!\!N}^{-1}{I\!\!R}^3 
\frac {\partial^2 {I\!\!K}}{\partial \bar Z^{\bar A} \partial Z^B} 
\left [D_{\bar \eta} \bar Z^{\bar A} D_{\eta} Z^B + D_{\bar \eta} Z^B 
D_{\eta}\bar Z^{\bar A} \right ], \, \nonumber
\end{eqnarray}  

\begin{eqnarray}
- \frac{2}{\kappa^3} {I\!\!R}^3 {\vert g(Z) \vert}^\alpha 
= - \frac{2}{\kappa^3} {I\!\!R}^3 \exp \left [\frac{\alpha}{2} ({I\!\!K} + 
\log {\vert g(Z) \vert}^2) \right ] = - \frac{2}{\kappa^3}{I\!\!R}^3 
\exp(\frac{\alpha}{2} G(Z) ).\label{cinco}
\end{eqnarray} 
The thirth term in the expression (\ref{dos}) is identical 
to the right term of (\ref{tres}) and they are cancelled in the action. On 
the other hand, the last two terms in (\ref{dos}) are a total derivative and 
may be ignored. In the expression (\ref{cinco}) we have introduced 
the superfunction $G(Z^A, \bar Z^{\bar A})$ as a special combination of the 
super-K\"ahler potential ${I\!\!K}(Z^A, \bar Z^{\bar A})$ and of the 
spatially homogeneous superpotential $g(Z)$

\begin{equation}
G(Z, \bar Z) = {I\!\!K}(Z, \bar Z) + log|g(Z)|^2,
\label{seis}
\end{equation}
the K\"ahler superfunction $G(Z^A, \bar Z^{\bar A})$ is invariant under the 
super-transformations

\begin{eqnarray}
g(Z) &\to& g(Z) \exp{f(Z)}, \nonumber\\
{I\!\!K}(Z, \bar Z) &\to& {I\!\!K}(Z, \bar Z) - f(Z) - \bar f(\bar Z),
\label{siete}
\end{eqnarray}
where the super-K\"ahler potential ${I\!\!K}(Z,\bar Z)$ is defined on the 
complex superfield $Z^A$ related to $\Phi(Z, \bar Z)$ (\ref{weyl}). 
We denoted the derivatives of the K\"ahler superfunction by 
$\frac{\partial G}{\partial z^A} = G,_A \equiv G_A$, 
$\frac{\partial G}{\partial {\bar z^{\bar A}}} 
= G,_{\bar A} \equiv G_{\bar A}$, 
$\frac{\partial^n G}{\partial z^A \partial z^B \partial {\bar z^{\bar C}}
 \cdots \partial {\bar z^{\bar D}}} = G,_{AB \bar C \cdots \bar D} 
\equiv G_{AB \bar C \cdots \bar D}$ and the 
K\"ahler supermetric is $G_{A \bar B}$ = $G_{\bar B A}$ = $K_{A \bar B}$, 
and their inverse  $G^{A \bar B}$ such as 
$G^{A \bar B}G_{\bar B D} = \delta^A_B$ can be used to define 
$G^A \equiv G^{A \bar B} \, G_{\bar B}$ and $G^{\bar B} \equiv 
G_A G^{A \bar B}$. 

So, the superfield action (\ref{uno}) becomes
\begin{eqnarray}
S &=& \int \left \{ -\frac{3}{\kappa^2} {I\!\!N}^{-1} {I\!\!R} D_{\bar\eta}
{I\!\!R} D_{\eta} {I\!\!R} + \frac{3}{\kappa^2} \sqrt{k} {I\!\!R}^2 - 
\frac{2}{\kappa^3}{I\!\!R}^3 e^{\frac{\alpha G}{2}} \right.\nonumber\\
&& +\left. \frac{1}{2\kappa^2}{I\!\!N}^{-1}{I\!\!R}^3 
G_{\bar A B}\left [ D_{\bar\eta} \bar Z^{\bar A}
D_{\eta} Z^B + D_{\bar\eta} Z^B D_{\eta} \bar Z^{\bar A}\right ]  
\right \} d \eta 
d \bar\eta dt \, .
\label{ocho}
\end{eqnarray}
Now, this action is determined only by terms of the one arbitrary K\"ahler
superfunction $G(Z^A, \bar Z^{\bar A})$. Perhaps, it is important to 
mention,
that in the supergravity theory \cite{diesisiete} it is also possible to 
introduce the Weyl transformations $\phi(z,\bar z) \to -\frac{3}{\kappa^2}
\exp(-\frac{\alpha}{3} K(z,\bar z))$ and the vierbein 
$e_\mu^a \to \exp(\frac{\alpha}{6}K(z,\bar z))$ with an arbitrary parameter
$\alpha$. However, the terms in the supergravity action can not be 
represented by the K\"ahler function $G(z,\bar z)$, which is due to the 
scalar curvature term, the kinetic term in the complex fields and 
auxiliary fields $A_\mu$ in the supergravity multiplet are eliminated 
only if $\alpha =1$ \cite{diesisiete}. 

The superfield action (\ref{ocho}) is invariant under the $n=2$ local 
conformal time supersymmetric transformations of $(t, \eta, \bar\eta)$. 
These transformations can be written as

\begin{eqnarray} 
\delta t &=& {I\!\! L}(t,\eta,\bar \eta) + \frac{1}{2}\bar \eta 
D_{\bar\eta}{I\!\!L}(t,\eta,\bar \eta) -\frac{1}{2}\eta D_{\eta} 
{I\!\!L}(t,\eta,\bar \eta),\label{nueve}\\
\delta \eta &=& \frac{i}{2}D_{\bar\eta} {I\!\!L}(t,\eta,\bar \eta), 
\qquad\qquad \delta \bar \eta = -\frac{i}{2} D_{\eta} {I\!\!L}
(t,\eta,\bar \eta), \nonumber
\end{eqnarray}
with the superfunction ${I\!\!L}(t,\eta,\bar\eta)$ defined in the 
superspace $(t, \eta, \bar\eta)$ as
\begin{equation}
{I\!\!L}(t,\eta,\bar\eta) = a(t) + i\eta \bar\beta^\prime(t) + i\bar\eta 
\beta^\prime(t) + b(t)\eta \bar\eta, 
\label{diez}
\end{equation}
where $D_{\eta} = \frac{\partial}{\partial \eta} + i\bar\eta 
\frac{\partial}{\partial t}$ and 
$D_{\bar\eta} = -\frac{\partial}{\partial \bar\eta} - 
i\eta\frac{\partial}{\partial t}$ are the supercovariant derivatives of 
the global conformal supersymmetry, which have dimension 
$[D_\eta]=l^{-1/2}$, $a(t)$ is a local time reparametrization parameter, 
$\beta^\prime(t) = N^{-1/2}\beta(t)$ is the Grassmann complex parameter
of the local conformal supersymmetric transformations 
(\ref{nueve}), and $b(t)$ is the parameter of the local $U(1)$ 
rotations on the complex $\eta$. 

In order to have the component action for 
(\ref{ocho}) we must expand in Taylor series the superfields
${I\!\!N}(t, \eta, \bar\eta)$, ${I\!\!R}(t, \eta, \bar\eta)$,
$Z^A(t, \eta, \bar\eta)$ and $\bar Z^{\bar A}(t, \eta, \bar\eta)$ with 
respect to $\eta, \bar\eta$. Due to the anticommuting properties of the 
$\eta, \bar\eta$ we see, that this series expansion ends up with the 
second term, as in the case of the superfunction in (\ref{diez}). For the 
one-dimensional gravity superfield ${I\!\!N}(t, \eta, \bar\eta)$ we have 
the following series expansion 
\begin{equation}
{I\!\!N}(t, \eta, \bar\eta) = N(t) + i\eta\bar\psi^\prime(t) + i\bar\eta 
\psi^\prime(t) + \eta \bar\eta V^\prime(t), 
\label{once}
\end{equation}
where we have introduced the reparametrization 
$\psi^\prime(t)= N^{1/2}\psi(t)$, $\bar\psi^\prime(t)= 
N^{1/2}\bar\psi(t)$ and $V^\prime(t)= NV + \bar\psi \psi$. 
The superfield (\ref{once}) transforms as one-dimensional vector field 
under the local conformal time supersymmetric transformations 
(\ref{nueve}). This transformation law may be written as
\begin{equation}
\delta {I\!\!N}=  ({I\!\!L}{I\!\!N}\dot ) + \frac{i}{2}D_{\bar\eta}{I\!\!L}
D_{\eta}{I\!\!N} + \frac{i}{2}D_{\eta}{I\!\!L}D_{\bar\eta}{I\!\!N}\, . 
\label{doce}
\end{equation}
The components of the superfield 
${I\!\!N}(t,\eta,\bar\eta)$ in (\ref{once}) are gauge fields of the 
one-dimensional $n=2$ extended supergravity. The transformations law for the 
components $N(t)$, $\psi(t)$, $\bar\psi(t)$ and $V(t)$ of the superfield 
(\ref{once}) may be obtained from (\ref{doce}) 

\begin{eqnarray}
\delta N &=& (aN)^. + \frac{i}{2}(\beta \bar\psi + \bar\beta \psi),
\qquad\qquad \delta \psi = (a \psi)^. + D\beta - \frac{i}{2} 
\hat b \psi, \label{trece}\\
\delta \bar\psi &=& (a \bar\psi)^. + D\bar\beta + \frac{i}{2} 
\hat b \bar\psi, \qquad\qquad \delta V = (aV)^. + \dot{\hat b}
, \, \nonumber \end{eqnarray}
where $D \beta = \dot \beta + \frac{i}{2} \beta V$ and $D \bar\beta = 
\dot {\bar\beta} - \frac{i}{2} \bar\beta V$ are the $U(1)$ covariant 
derivatives and $\hat b = b - \frac{1}{2N} (\beta \bar\psi - 
\bar\beta \psi)$. 

The Taylor series expansion for the superfield 
${I\!\!R}(t, \eta, \bar\eta)$ has the form
\begin{equation}
{I\!\!R}(t, \eta, \bar\eta)= R(t) + i\eta\bar\lambda^\prime(t) + 
i\bar\eta \lambda^\prime(t) + \eta \bar\eta B^\prime(t), 
\label{catorce}
\end{equation}
where $\lambda^\prime(t)= \kappa N^{1/2}\lambda(t),$ 
$\bar\lambda^\prime(t)= \kappa N^{1/2}\bar\lambda(t)$ 
and $B^\prime(t)= \kappa NB + \frac{\kappa}{2}
(\bar\psi\lambda - \psi \bar\lambda)$. The transformation rule for the real 
scalar superfield ${I\!\!R}(t, \eta, \bar\eta)$ under supersymmetric 
transformation is  
\begin{equation}
\delta {I\!\!R}= {I\!\!L}\dot{I\!\!R}+ \frac{i}{2} D_{\bar\eta}{I\!\!L}
D_{\eta}{I\!\!R} + \frac{i}{2}D_{\eta}{I\!\!L}D_{\bar\eta}{I\!\!R}. 
\label{quince}
\end{equation}
The component $B(t)$ in (\ref{catorce}) is an auxiliary degree of 
freedom (nondynamical variable), $\lambda(t)$ and $\bar\lambda(t)$ are 
the fermionic superpartners of the scale factor $R(t)$. Their 
transformations law has the form
\begin{eqnarray}
\delta R &=& a\dot R + \frac{i\kappa}{2}(\beta \bar\lambda + 
\bar\beta \lambda),\qquad\qquad  \delta \lambda = a \dot\lambda + 
\frac{\beta}{2\kappa} \left [\frac{DR}{N} + i\kappa B \right ] - 
\frac{i}{2} \hat b \lambda, \label{diesiseis}\\
\delta \bar\lambda &=& a \dot{\bar\lambda} + \frac{\bar \beta}{2\kappa}
\left [\frac{DR}{N} - i\kappa B \right ] + \frac{i}{2} \hat b \bar\lambda, 
\qquad\qquad
\delta B = a\dot B + \frac{1}{2N}(\bar\beta \tilde D \lambda - 
\beta \tilde D\bar\lambda),\, \nonumber 
\end{eqnarray}     
where $DR = \dot R - \frac{i\kappa}{2}(\psi \bar\lambda + 
\bar\psi \lambda)$,
$\tilde D \lambda = D\lambda - \frac{i}{2\kappa} \left [\frac{DR}{N} + 
i\kappa B \right ]\psi$ and $\tilde D \bar\lambda = D \bar\lambda - 
\frac{i}{2\kappa} \left [\frac{DR}{N} - i\kappa B \right ] \bar\psi $ are 
the supercovariant derivatives, and $D\lambda = \dot\lambda + 
\frac{i}{2} V \lambda $ and $D\bar\lambda = \dot{\bar\lambda} - 
\frac{i}{2} V \bar\lambda$ are the $U(1)$ covariant derivatives.    

The spatially homogeneous complex scalar matter superfields 
$Z^A (t, \eta, \bar\eta)$ and $\bar Z^{\bar A}(t, \eta, \bar \eta) = 
(Z^A)^\ast$ consist of a set of spatially homogeneous matter fields 
$z^A(t)$ and $\bar z^{\bar A}(t)$ (A= 1,2,..,n), four fermionic degrees of 
freedom $\chi^A(t)$, $\bar\chi^{\bar A}(t)$, $\phi^A(t)$ and 
$\bar \phi^{\bar A}(t)$, as well as bosonic auxiliary fields $F^A(t)$ and 
$\bar F^{\bar A}(t)$.

The components of the matter superfields $ Z^A(t, \eta, \bar\eta) $ and 
$\bar Z^{\bar A}(t, \eta, \bar\eta)$ may be written as
\begin{equation}
 Z^A(t, \eta, \bar\eta) = z^A(t) + i\eta \chi^{\prime A}(t)+ 
i\bar\eta\phi^{\prime A}(t) + \eta \bar\eta F^{\prime A}(t), 
\label{diesisiete}
\end{equation} 

\begin{equation}
 \bar Z^{\bar A}(t, \eta, \bar\eta)= \bar z^{\bar A}(t) + 
i\eta \bar\phi^{\prime \bar A}(t) + i\bar\eta \bar\chi^{\prime \bar A}(t) + 
\eta \bar\eta \bar F^{\prime \bar A}(t), 
\label{diesiocho}
\end{equation}
where $\chi^{\prime A}(t) = N^{1/2} \chi^A(t)$, $\phi^{\prime A}(t)= 
N^{1/2} \phi^A(t)$, $ F^{\prime A}(t)= NF^A - \frac{1}{2}(\psi \chi^A - 
\bar\psi \phi^A)$ and $ \bar F^{\prime \bar A}(t)= N \bar F^{\bar A} - 
\frac{1}{2} (\psi \bar \phi^{\bar A} - \bar\psi \bar\chi^{\bar A})$

The transformation rule for the superfields $Z^A(t, \eta, \bar\eta)$ and
$\bar Z^{\bar A}(t, \eta, \bar\eta)$ may be written as  
\begin{eqnarray}  
\delta Z^A &=& {I\!\!L} \dot Z^A + \frac{i}{2} D_{\bar\eta} 
{I\!\!L} D_\eta
Z^A + \frac{i}{2}D_\eta {I\!\!L} D_{\bar\eta} Z^A \, , \label{diesinueve}\\
\delta \bar Z^{\bar A} &=& {I\!\!L} \dot {\bar Z}^{\bar A} + 
\frac{i}{2} D_\eta {I\!\!L}D_{\bar\eta} \bar Z^{\bar A} + 
\frac{i}{2} D_{\bar\eta} {I\!\!L} D_\eta \bar Z^{\bar A}\, . \label{veinte}
\end{eqnarray} 
From this superfield transformations we can obtain easily the  
transformations law for the components of the matter superfields 
$Z^A(t, \eta, \bar\eta)$ and $\bar Z^{\bar A}(t, \eta, \bar\eta)$. We get

\begin{eqnarray}
\delta z^A &=& a \dot z^A + \frac{i}{2}(\beta \chi^A + 
\bar\beta \phi^A),
\qquad\qquad\qquad\qquad \delta \bar z^{\bar A} = a \dot{\bar z}^{\bar A} + 
\frac{i}{2}(\bar\beta \bar\chi^{\bar A} + \beta \bar\phi^{\bar A}), 
\nonumber\\
\delta \chi^a &=& a \dot \chi^A + \frac{i}{2} \hat b \chi^A +
\frac{\bar \beta}{2} \left [\frac{D z^A}{N} - iF^A \right ], \qquad
\delta \bar\chi^{\bar A} = a \dot{\bar\chi}^{\bar A} - \frac{i}{2} \hat b
\bar \chi^{\bar A} + \frac{\beta}{2} \left [\frac{D {\bar z}^{\bar A}}{N} + 
i\bar F^{\bar A} \right ], \label{v-1} \\
\delta \phi^A &=& a \dot \phi^A - \frac{i}{2} \hat b \phi^A + 
\frac{\beta}{2} \left [\frac{Dz^A}{N} + iF^A \right ], \qquad
\delta \bar\phi^{\bar A} = a \dot{\bar \phi}^{\bar A} + \frac{i}{2} \hat b 
\bar\phi^{\bar A} + \frac{\bar\beta}{2} 
\left [\frac{D\bar z^{\bar A}}{N} - i\bar F^{\bar A} \right ],\nonumber\\
\delta F^A &=& a \dot F^A + \frac{1}{2N} \left [\bar\beta 
\hat D \phi^A - \beta \hat D \chi^A \right ], \qquad\qquad
\delta \bar F^{\bar A} = a \dot{\bar F}^{\bar A} + \frac{1}{2N} 
\left [\bar\beta  \hat D \bar\chi^{\bar A} - \beta 
\hat D \bar\phi^{\bar A} \right ], \nonumber 
\end{eqnarray} 
where $Dz^A = \dot z^A - \frac{i}{2}(\bar\psi \phi^A + 
\psi \chi^A)$, $D\bar z^{\bar A} = \dot {\bar z}^{\bar A} - 
\frac{i}{2} (\psi \bar\phi^{\bar A} + \bar\psi \bar\chi^{\bar A})$, 
$\hat D \phi^A = D\phi^A - \frac{1}{2}(N^{-1} Dz^A + iF^A)\psi$, 
$\hat D \bar \phi^{\bar A} = D \bar \phi^{\bar A} - \frac{1}{2}(N^{-1}
D \bar z^{\bar A} - i \bar F^{\bar A})\bar \psi$, 
$\hat D \chi^A = D\chi^A - \frac{1}{2}(N^{-1} Dz^A - iF^A)\psi$, 
and $\hat D \bar \chi^{\bar A} = D \bar \chi^{\bar A} - \frac{1}{2}(N^{-1}
D \bar z^{\bar A} + i \bar F^{\bar A})\psi$ are the supercovariant 
derivatives,
and $D\phi^A = \dot \phi^A + \frac{i}{2} V \phi^A$,
$D\bar \phi^{\bar A} = \dot{\bar \phi}^{\bar A} - \frac{i}{2} V 
\bar \phi^{\bar A}$, $D\chi^A = \dot \chi^A - \frac{i}{2} V \chi^A$ and
$D\bar \chi^{\bar A} = \dot{\bar \chi}^{\bar A} + \frac{i}{2} V 
\bar \chi^{\bar A}$ are the $U(1)$ covariant derivatives.

\section {Supersymmetric Lagrangian and susy Breaking}

It is clear, that the superfield action (\ref{ocho}) is invariant under
the $n=2$ local conformal time supersymmetry. We can write the expression,
which is found under the integral (\ref{ocho}) by means of certain
superfunction $f({I\!\!N}, {I\!\!R}, Z, \bar Z)$. Then, the 
infinitesimal small transformations of the action (\ref{ocho}) under
the superfield transformations 
(\ref{doce},\ref{quince},\ref{diesinueve},\ref{v-1}) have the form

\begin{equation}
\delta S = \frac{i}{2}\int \{D_{\bar \eta}({I\!\!L} D_{\eta}
f({I\!\!N},{I\!\!R}, Z, \bar Z)) +  
D_{\eta}({I\!\!L} D_{\bar \eta}f({I\!\!N},{I\!\!R}, Z, \bar Z)) \}
d\eta d\bar\eta dt, \label {v-2}
\end{equation}     
we can see, that under the integration it gives a total derivative, 
{\it i.e} the action (\ref{ocho}) is invariant under the superfield 
transformations (\ref{doce},\ref{quince},\ref{diesinueve},\ref{v-1}).

For the K\"ahler superfunction in (\ref{ocho}) we have the following 
Taylor series expansion
\begin{eqnarray}
G(Z, \bar Z)&=& G(z,\bar z) + G_A(z, \bar z)(Z^A - z^A) + 
G_{\bar A}(z, \bar z)(\bar Z^{\bar A} - \bar z^{\bar A}) \nonumber\\
& &+ \frac{1}{2} G_{AB}(z, \bar z) (Z^A - z^A)(Z^B - z^B)+
\frac{1}{2} G_{\bar A \bar B}(z, \bar z)(\bar Z^{\bar A} - 
\bar z^{\bar A}) (\bar Z^{\bar B} - \bar z^{\bar B}) \nonumber\\
&&+ G_{\bar A B}(z, \bar z)(\bar Z^{\bar A} - \bar z^{\bar A})(Z^B - z^B), 
\label{v-3}
\end{eqnarray}
where the first term in the expansion is the ordinary K\"ahler function 
of the supergravity theories interacting with complex scalar matter 
supermultiplets \cite{diesisiete}. So, making the corresponding 
operations in the superfield action (\ref{ocho}), and in order to have 
the correct kinetic term for the ``fermionic fields", we make the 
following redefinition of the fields 
$\lambda = \frac{\lambda}{3\sqrt R}$, $\chi^A = \frac{\chi^A}{\sqrt R^3}$ 
and $\phi^A = \frac{\phi^A}{\sqrt R^3}$.

After integration over $\eta, \bar\eta$ the superfield action (\ref{ocho})
may be written in the usual form $S = \int L dt$, the lagrangian $L$
contains terms with auxiliary fields $B(t)$, $F(t)$ and $\bar F(t)$ of 
superfields ${I\!\!R}$, $Z$ and $\bar Z$ respectively. We can write the 
lagrangian as the sum $L = \tilde L + L_{aux}$ of lagrangian without 
auxiliary fields and lagrangian for the auxiliary fields respectively. 
Explicitly we have 
\begin{eqnarray}
\tilde L &=& - \frac{3}{N\kappa^2} R(DR)^2 + \frac{2i}{3}\bar\lambda 
D\lambda + \frac{\sqrt k \sqrt R}{\kappa} (\bar\psi \lambda - 
\psi \bar\lambda) + \frac{2N \sqrt k}{3R} \bar\lambda \lambda + 
\frac{R^3}{N\kappa^2}G_{\bar A B}D\bar z^{\bar A} Dz^B \nonumber \\
&+& \frac{i}{2\kappa}Dz^B(\bar\lambda G_{\bar A B} 
\bar\chi^{\bar A} + \lambda G_{\bar A B} \bar\phi^{\bar A}) + 
\frac{i}{2\kappa}D\bar z^{\bar A}(\bar\lambda G_{\bar A B} \phi^B + 
\lambda G_{\bar A B}\chi^B)-\frac{i}{\kappa^2}G_{\bar A B} 
\bar\chi^{\bar A} \tilde D \chi^B \nonumber \\
&-& \frac{i}{\kappa^2}G_{\bar A B} \bar\phi^{\bar A} \tilde D \phi^B - 
\frac{N}{\kappa^2 R^3} G_{\bar A B \bar C D} \bar\chi^{\bar A} \chi^B
\bar\phi^{\bar C} \phi^D - \frac{1}{4\kappa \sqrt R^3} (\psi \bar\lambda - 
\bar\psi \lambda)G_{\bar A B}(\bar\chi^{\bar A} \chi^B + 
\phi^B \bar\phi^{\bar A}) \nonumber \\
&+& \frac{N}{2\kappa R^3}(\bar\lambda 
G_{\bar A B \bar C}\bar\chi^{\bar C} - \lambda G_{\bar A B C}
\chi^C)\phi^B \bar\phi^{\bar A} + \frac{N}{2\kappa R^3}(\bar\lambda 
G_{\bar A B C} \phi^C - \lambda G_{\bar A B \bar C}\bar\phi^{\bar C}) 
\bar\chi^{\bar A} \chi^B\nonumber \\
&+& \frac{N}{3R^3} G_{\bar A B}(\bar\chi^{\bar A} \chi^B + 
\phi^B \bar\phi^{\bar A})\bar\lambda \lambda - \frac{4N}{3\kappa} 
e^{\frac{\alpha G}{2}} \bar\lambda \lambda - 2\frac{N}{\kappa^3}
(e^{\frac{\alpha G}{2}})_{,A B}\chi^A \phi^B  \label{v-4} \\
&-& 2\frac{N}{\kappa^3}(e^{\frac{\alpha G}{2}})_{,\bar A \bar B}
\bar\phi^{\bar A} \bar\chi^{\bar B} - 
\frac{2N}{\kappa^3}(e^{\frac{\alpha G}{2}})_{,\bar A B} 
(\bar\chi^{\bar A}\chi^B + \phi^B \bar\phi^{\bar A}) - 
\frac{2N}{\kappa^2} \bar\lambda \left [(e^{\frac{\alpha G}{2}})_A 
\phi^A + (e^{\frac{\alpha G}{2}})_{\bar A} \bar\chi^{\bar A} \right] 
\nonumber \\
&+& \frac{2N}{\kappa^2}\lambda \left [(e^{\frac{\alpha G}{2}})_A \chi^A + 
(e^{\frac{\alpha G}{2}})_{\bar A} \bar\phi^{\bar A} \right ]   
+ \frac{\sqrt R^3}{\kappa^3} (e^{\frac{\alpha G}{2}})_{, A} 
(\psi \bar\chi^{\bar A} - \bar\psi \phi^A) 
+ \frac{\sqrt R^3}{\kappa^3} (e^{\frac{\alpha G}{2}})_{,\bar A} 
(\psi \bar\phi^{\bar A} - \bar\psi \bar\chi^{\bar A})\nonumber \\
&-& \frac{\sqrt R^3}{\kappa^2}(\bar\psi \lambda - \psi \bar\lambda) 
e^{\frac{\alpha G}{2}}, \nonumber 
\end{eqnarray} 

and the lagrangian for the auxiliary fields has the form
\begin{equation}
L_{aux} = L_{aux~~FRW} + L_{aux~~kinetical~~term} + 
L_{aux~~potential~~term} = 
L_{aux~~B} + L_{aux~~F,\bar F};
\label{v-5}
\end{equation}
explicitly the lagrangian for the auxiliary field $B$  

\begin{equation}
N^{-1} L_{aux~B}= -3RB^2 + \left [-\frac{\kappa}{3R} \bar\lambda \lambda
+ \frac{6\sqrt k}{\kappa}R + \frac{3}{2\kappa R} G_{\bar A B}
(\bar\chi^{\bar A}\chi^B + \phi^B \bar\phi^{\bar A}) 
- \frac{6}{\kappa^2} R^2 e^{\frac{\alpha G}{2}} \right ]\, ,
\label{v-6}
\end{equation}
and the lagrangian for the auxiliary fields $F, \bar F$ is

\begin{eqnarray}
N^{-1} L_{aux~~F, \bar F} &=& F^B \left [\frac{1}{2\kappa} G_{\bar A B} 
(\bar\lambda \bar\chi^{\bar A}
- \lambda \bar\phi^{\bar A}) + \frac{1}{\kappa^2} G_{\bar A B \bar C}
\bar\phi^{\bar C} \bar\chi^{\bar A} - \frac{2R^3}{\kappa^3}
(e^{\frac{\alpha G}{2}})_{, B} \right ] \label{v-7} \\
&+& \bar F^{\bar A} \left [\frac{1}{2\kappa} G_{\bar A B}
(\bar\lambda \phi^B - \lambda \chi^B) + \frac{1}{\kappa^2} 
G_{\bar A B C} \chi^C \phi^B - \frac{2R^3}{\kappa^3}(e^{\frac{\alpha G}
{2}})_{, \bar A} \right ] +\frac{R^3}{\kappa^2} G_{\bar A B} 
\bar F^{\bar A} F^B. \nonumber  
\end{eqnarray}    

The equations for the auxiliary fields $B$, $F^A$ and $\bar F^{\bar A}$ are
algebraical and may be determined from the component action of
(\ref{ocho}) by taking the variation with respect to them. We get the 
following solutions for the auxiliary fields
\begin{eqnarray}
B&=& -\frac{\kappa}{18 R^2} \bar\lambda \lambda + \frac{\sqrt k}{\kappa} +
\frac{1}{4\kappa R^2} G_{\bar A B}(\bar\chi^{\bar A} \chi^B + 
\phi^B \bar\phi^{\bar A}) - \frac{R}{\kappa^2} e^{\frac{\alpha G}{2}},
\nonumber\\     
 F^D &=& -\frac{\kappa}{2 R^3} (\bar\lambda \phi^D - \lambda \chi^D) -
\frac{1}{R^3} G^{D \bar A} G_{\bar A B C} \chi^C \phi^B + 
\frac{2}{\kappa} G^{D \bar A} (e^{\frac{\alpha G}{2}}),_{\bar A}, 
\label{v-8}\\
\bar F^{\bar D} &=& -\frac{\kappa}{2 R^3} (\bar\lambda \chi^{\bar D} - 
\lambda \bar\phi^{\bar D}) - \frac{1}{R^3} G^{\bar D B} G_{\bar A B \bar C}
\bar\phi^{\bar C} \bar\chi^{\bar A} + 
\frac{2}{\kappa} G^{\bar D B} (e^{\frac{\alpha G}{2}}),_B ,\, \nonumber
\end{eqnarray}
where $(~~~)_{,A}$ and $(~~~)_{,\bar A}$ are derivatives with respect to
$z^A$ and $\bar z^{\bar A}$.
After substituting them into the component action we have the total 
supersymmetric action $\int L dt = \int(\tilde L + L_{aux~~B} + 
L_{aux~~F, \bar F})dt$. We get the following expression for the total 
Lagrangian
\begin{eqnarray}
L &=&  - \frac{3}{\kappa^2} \frac{R (DR)^2}{N} - N R^3 U(R,z,\bar z) + 
\frac{2i}{3} \bar\lambda D\lambda + \frac{N \sqrt k}{3R} \bar\lambda 
\lambda - \frac{N}{\kappa} e^{\frac{\alpha G}{2}} \bar\lambda \lambda 
\nonumber\\
&+& \frac{\sqrt k}{\kappa}\sqrt{R} \left (\bar\psi \lambda - 
\psi \bar\lambda \right) + \frac{R^3}{N\kappa^2}G_{\bar A B} 
D\bar z^{\bar A} Dz^B  + \frac{i}{2\kappa} Dz^B \left(\bar\lambda 
G_{\bar A B} \bar\chi^{\bar A} + \lambda G_{\bar A B} 
\bar\phi^{\bar A}\right ) \nonumber\\
&+& \frac{i}{2\kappa} D\bar z^{\bar A} \left( \bar\lambda G_{\bar A B}
\phi^B+\lambda G_{\bar A B} \chi^B \right) - \frac{i}{\kappa^2} 
G_{\bar A B} \left( \bar\chi^{\bar A} \tilde D \chi^B + 
\bar\phi^{\bar A} \tilde D \phi^B \right ) \nonumber\\
&-& \frac{N}{\kappa^2 R^3} R_{\bar A B \bar C D} \bar\chi^{\bar A} 
\chi^B \bar\phi^{\bar C} \phi^D - \frac{i}{4\kappa \sqrt{R^3}}
\left( \psi \bar \lambda - \bar \psi \lambda \right ) G_{\bar AB} 
\left( \bar \chi^{\bar A} \chi^B + \phi^B \bar\phi^{\bar A} 
\right) \nonumber\\
&+& \frac{3N}{16\kappa^2 R^3} \lbrack G_{\bar A B} 
\left (\bar\chi^{\bar A} \chi^B + \phi^B \bar\phi^{\bar A}\right) 
\rbrack^2 + \frac{3 \sqrt k}{2 \kappa^2 R}G_{\bar A B}\left 
(\bar\chi^{\bar A} \chi^B + \phi^B \bar\phi^{\bar A}\right ) 
\label{v-9} \\
&-& \frac{3N}{2\kappa^3} e^{\frac{\alpha G}{2}} G_{\bar A B} 
\left( \bar\chi^{\bar A} \chi^B + \phi^B \bar\phi^{\bar A}\right )
- \frac{2N}{\kappa^3} (e^{\frac{\alpha G}{2}})_{, A B}
\chi^A \phi^B - \frac{2}{\kappa^3} 
N (e^{\frac{\alpha G}{2}})_{,\bar A \bar B}
\bar\phi^{\bar A} \bar\chi^{\bar B} \nonumber\\
&-& \frac{2N}{\kappa^3}(e^{\frac{\alpha G}{2}})_{,\bar A B}
\left ( \bar\chi^{\bar A} \chi^B + \phi^B \bar\phi^{\bar A}\right ) 
- \frac{N}{\kappa^2} \bar\lambda \lbrack (e^{\frac{\alpha G}{2}})_{, A} 
\phi^A + (e^{\frac{\alpha G}{2}})_{,\bar A}
\bar\chi^{\bar A}\rbrack \nonumber\\
&+& \frac{N}{\kappa^2} \lambda \lbrack (e^{\frac{\alpha G}{2}})_{, A} 
\chi^A + (e^{\frac{\alpha G}{2}})_{,\bar A} \bar\phi^{\bar A}\rbrack
- \frac{\sqrt{R^3}}{\kappa^2} \left( \bar\psi \lambda - \psi \bar\lambda
\right ) e^{\frac{\alpha G}{2}} \nonumber\\
&+& \frac{\sqrt{R^3}}{\kappa^3}(e^{\frac{\alpha G}{2}})_{, A}
\left ( \psi \chi^A - \bar\psi \phi^A \right )  + 
\frac{\sqrt{R^3}}{\kappa^3} (e^{\frac{\alpha G}{2}})_{,\bar A} 
\left (\psi \bar \phi^{\bar A} - \bar \psi \bar \chi^{\bar A}\right ), 
 \nonumber
\end{eqnarray}
where $DR$, $Dz^A$, $D\chi^B$, $D\phi^B$ and $D\lambda$ are defined before
$\tilde D \chi^B = D \chi^B + \Gamma^B_{CD} \dot {z}^C \chi^D$,
$\tilde D \phi^B = D\phi^B + \Gamma^B_{CD} \dot {z}^C \phi^D$,
and $R_{\bar A B \bar C D}$ is the curvature tensor of the K\"ahler manifold 
defined by the coordinates $z^A, \bar z^{\bar B}$ with the metric 
$G_{A \bar B}$, and $\Gamma^B_{CD}=G^{B \bar A}\, G_{\bar ACD}$ are the 
Christoffel symbols in the definition of covariant derivative and their 
complex conjugate. After elimination of the auxiliary fields the Lagrangian 
(\ref{v-9}) only depends on covariant magnitudes, as we will show, it is
a specific classical Lagrangian of supersymmetric quantum mechanics.  
In the Lagrangian (\ref{v-9}) the potential term is written as
\begin{equation}
U(R, z, \bar z)= - \frac{3k}{\kappa^2 R^2} + \frac{6\sqrt k}{\kappa^3 R}
e^{\frac{\alpha G}{2}} + V_{eff}(z, \bar z),
\label{v-10}
\end{equation}
where the effective potential of the scalar matter fields is
\begin{equation}
V_{eff} = \frac{4}{\kappa^4} \left [(e^{\frac{\alpha G}{2}})_{\bar A} 
G^{\bar A D} (e^{\frac{\alpha G}{2}})_D - \frac{3}{4} e^{\alpha G} \right ]
= \frac{e^{\alpha G}}{\kappa^4} \left [\alpha^2 G^A G_A - 3 \right ].
\label{v-11}
\end{equation}   

In order to discuss the implication of spontaneous supersymmetry breaking 
we need to display the potential (\ref{v-10}) in terms of the 
auxiliary fields
\begin{equation}
U(R, z, \bar z) = \frac{\bar F^{\bar A}G_{\bar A B}F^B}{\kappa^2} - 
\frac{3B^2}{R^2}, \label{v-12}
\end{equation}
where the bosonic terms (\ref{v-8}) are 

\begin{equation}
F^A = \frac{\alpha}{\kappa}e^{\frac{\alpha G(z \bar z)}{2}} 
G^A(z, \bar z),
\label{v-13}
\end{equation}

\begin{equation} 
B = \frac{\sqrt k}{\kappa} - 
\frac{R}{\kappa^2}e^{\frac{\alpha G(z, \bar z)}{2}}. \label{v-14}
\end{equation}
The supersymmetry is spontaneous breaking if the auxiliary fields 
(\ref{v-13}) of the matter supermultiplets get non-vanishing vacuum 
expectation values. According to our assumption at the minimum in 
(\ref{v-11}) for $k=0$,  $V_{eff}(z_o, \bar z_0) = 0$, but 
$<F^A> = F^A(z_0, \bar z_0)\not = 0$ and, thus, the supersymmetry is 
broken.

The scalar field potential (\ref{v-11}) consists of two terms, one of 
them is the potential for the scalar fields in the case of the global 
supersymmetry. Unlike the standard supersymmetric quantum mechanics the
potential is not positive-definite and allows spontaneous breaking of 
supersymmetry with vanishing classical vacuum energy. In this case the 
condition of the function $G(z, \bar z)$ must be fulfilled in the minimum 
of the potential $<G> = G(z_0, \bar z_0)$  
\begin{equation}
\frac{\partial V(z, \bar z)}{\partial z^A} \vert_{ z^A = z^A_0} = 0,
\qquad\qquad <G^A><G_A> = \frac{3}{\alpha^2},
\label{v-15}
\end{equation}
where $<G^A> = G^A(z_0, \bar z_0)$.

Due to the Lagrangian form (\ref{v-9}) in the simplest case $k=0$, the 
fermion bilineal terms $\lambda, \bar\lambda, \chi^A, \bar\chi^{\bar A},
\phi^A$ and $\bar \phi^{\bar A}$ are also the mass terms.

If supersymmetry is broken the fermion fields $\lambda, \bar\lambda$ will
be described by

\begin{equation}
\tilde \lambda = \lambda + \frac{1}{\kappa}e^{-\frac{\alpha <G>}{2}} 
\tilde \eta, \qquad\qquad  \bar {\tilde\lambda} = \bar \lambda + 
\frac{1}{\kappa}e^{-\frac{\alpha <G>}{2}} \tilde {\bar\eta},        
\label{v-16}
\end{equation}
where
\begin{equation}
\tilde \eta = <(e^{\frac{\alpha G}{2}})_{,A}>\phi^A + 
<(e^{\frac{\alpha G}{2}})_{,\bar A}>\chi^{\bar A},\qquad\qquad  
\bar {\tilde\eta} = <(e^{\frac{\alpha G}{2}})_{,A}>\chi^A + 
<(e^{\frac{\alpha G}{2}})_{,\bar A}>\phi^{\bar A}.
\label{v-17}
\end{equation}

After substitution the equations (\ref{v-16}) and (\ref{v-17}) into the mass 
term of Lagrangian (\ref{v-9}) has the following form
\begin{eqnarray}
L_{\it fermion~~mass~~term} &=& - 
N [\frac{1}{\kappa}e^{\frac{\alpha <G>}{2}} 
\bar\lambda \lambda + m_{AB}\chi^A \phi^B + m_{\bar A \bar B} 
\bar\phi^{\bar A}\bar\chi^{\bar B} \label{v-18}\\
&+& m_{\bar A B}(\bar\chi^{\bar A} \chi^B +
\phi^B \bar\phi^{\bar A})], \nonumber 
\end{eqnarray}
where $m_{AB}(z_0, \bar z_0), m_{\bar A \bar B}(z_0, \bar z_0)$ and
$m_{\bar A B}(z_0, \bar z_0)$ are the mass matrices depending on
$z^A_0 = <z^A(t)>$ vacuum expectation values of the scalar fields in the
minimum of the potential $V_{eff} (z_0, \bar z_0) = 0$. 

For the mass matrices we have the representation
\begin{eqnarray}
m_{AB}&=& \frac{ e^{\frac{\alpha <G>}{2}}}{\kappa^3} 
[\frac{\alpha^2}{4}<G_A>
<G_B> + \alpha <G_{AB}>], \nonumber\\ 
m_{\bar A \bar B} &=& \frac{e^{\frac{\alpha <G>}{2}}}{\kappa^3} 
[\frac{\alpha^2}{4}<G_{\bar A}><G_{\bar B}> + \alpha <G_{\bar A \bar B}>], 
\label{v-19}\\  
m_{\bar A B} &=& \frac{e^{\frac{\alpha <G>}{2}}}{\kappa^3} 
[\frac{\alpha^2}{4}<G_{\bar A}><G_B> + (\alpha +\frac{3}{2}) 
<G_{\bar A B}>]. \, \nonumber
\end{eqnarray} 

The ordering parameter of spontaneous breaking of supersymmetry is given by
the coefficient of $\bar\lambda \lambda$ term. From (\ref{v-18}) we identify
\begin{equation}
\frac{1}{\kappa}e^{\frac{\alpha <G>}{2}} = m_{3/2},
\label{vv}
\end{equation}
as the gravitino mass in the effective supergravity theories 
\cite{diesisiete}. 
 
\section {Canonical Formulation and the Constraints}

Now, we will proceed with the Hamiltonian analysis of this system. For this 
purpose we need to write the momenta conjugate to dynamical variables $R(t)$,
$z^A(t)$ and $\bar z^{\bar A}(t)$
\begin{eqnarray}
\pi_R &=& - \frac{6}{N\kappa^2} R(DR) \nonumber\\
\pi_A(z) &=& \frac{R^3}{N\kappa^2} G_{\bar B A} D\bar z^{\bar B} + 
\frac{i}{2\kappa}(\bar\lambda G_{\bar B A} \bar\chi^{\bar B} + 
\lambda G_{\bar B A} \bar\phi^{\bar B}) - \frac{i}{2\kappa^2} G_{\bar M B}
(\Gamma_{A D}^B \bar\chi^{\bar M}\chi^D + \Gamma_{A D}^B \bar\phi^{\bar M}
\phi^D), \nonumber \\
&=& p_A(\bar z) - \frac{i}{2\kappa^2} G_{\bar M B}(\Gamma_{A D} ^B 
\bar\chi^{\bar M}\chi^D + \Gamma_{A D}^B \bar\phi^{\bar M}\phi^D) 
\label{vv-1} \\
\bar\pi_{\bar A}(z) &=& \frac{R^3}{N\kappa^2} G_{\bar A B} Dz^B +
\frac{i}{2\kappa}(\bar\lambda G_{\bar A B} \phi^B + \lambda G_{\bar A B}
\chi^B) + \frac{i}{2\kappa^2} G_{\bar M B}(\Gamma_{\bar A \bar D} ^{\bar M}
\bar \chi^{\bar D} \chi^B + \Gamma_{\bar A \bar D} ^{\bar M} 
\bar \phi^{\bar D} \phi^B) \nonumber\\
&=& \bar p_{\bar A}(z) + \frac{i}{2\kappa^2} G_{\bar M B}
(\Gamma_{\bar A \bar D} ^{\bar M} \bar \chi^{\bar D}\chi^B + 
\Gamma_{\bar A \bar D} ^{\bar M} \bar \phi^{\bar D} \phi^B), \nonumber 
\end{eqnarray}
where $p_A(\bar z)$ and $p_{\bar A}(z)$ are the covariant momenta. With 
respect to the canonical Poisson brackets we have

\begin{eqnarray}
{\lbrace R, \pi_R \rbrace}= -1,\qquad\qquad {\lbrace z^B, \pi_A \rbrace}=
- \delta_A^B, \qquad\qquad {\lbrace \bar z^{\bar B}, \bar\pi_{\bar A} 
\rbrace} = - \delta_{\bar A}^{\bar B}. \label{vv-2} 
\end{eqnarray}

For the dynamical Grassmann variables $\lambda(t)$, $\chi(t)$ and $\phi(t)$
we have the following constraints
\begin{eqnarray}
\Pi_\lambda &\equiv& \pi_\lambda + \frac{i}{3} \bar\lambda \approx 0, 
\qquad\qquad\qquad\qquad\qquad\qquad \Pi_{\bar\lambda} \equiv 
\pi_{\bar \lambda} + \frac{i}{3}\lambda \approx 0,
\nonumber\\
\Pi_A(\chi) &\equiv& \pi_A(\chi) - \frac{i}{2\kappa^2} G_{\bar B A} 
\bar \chi^{\bar B} \approx 0, \qquad\qquad\qquad \Pi_{\bar A}(\chi) \equiv 
\pi_{\bar A}(\chi) - \frac{i}{2\kappa^2} G_{\bar A B}\chi^B \approx 0 ,
\label{vv-3} \\
\Pi_A(\phi) &\equiv& \pi_A(\phi) - \frac{i}{2\kappa^2} G_{\bar B A}
\bar\phi^{\bar B} \approx 0, \qquad\qquad\qquad \Pi_{\bar A}(\phi) \equiv 
\pi_{\bar A}(\phi) - \frac{i}{2\kappa^2} G_{\bar A B}\chi^B \approx 0, 
\nonumber 
\end{eqnarray}
where $\pi_\lambda = \frac{dL}{d \dot\lambda}$, $\pi_A(\chi) = 
\frac{dL}{d \dot\chi^A}$ and $\pi_A(\phi ) = \frac{dL}{d \dot\phi^A}$ 
are the momenta conjugate to the anticommuting variables 
$\lambda(t)$, $\chi(t)$ and $\phi(t)$ respectively. The constraints 
(\ref{vv-3}) are of the second class, and, therefore, they can be 
eliminated by the Dirac procedure. We define the matrix constraints 
\begin{eqnarray}
C_{\lambda \bar\lambda} = \frac{2}{3}i, \qquad\qquad C_{\bar A B}(\chi)=
- \frac{i}{\kappa^2} G_{\bar A B}, \qquad\qquad C_{\bar A B}(\phi)=
- \frac{i}{\kappa^2} G_{\bar A B},
\label{vv-4}
\end{eqnarray}
and their inverse matrices as
\begin{eqnarray}
C^{\lambda \bar\lambda} = (C_{\lambda \bar\lambda})^{-1}= - \frac{3}{2}i,
\qquad\qquad C^{\bar A B}(\chi) = i\kappa^2 G^{\bar A B}, \qquad\qquad
C^{\bar A B}(\phi) = i\kappa^2 G^{\bar A B}.
\label{vv-5}
\end{eqnarray}
The Dirac brackets ${\lbrace , \rbrace}^\ast$ are then defined by 
\begin{eqnarray}
\lbrace f,g\rbrace^\ast = \lbrace f,g\rbrace - \lbrace f, \Pi_i\rbrace
(C^{-1})^{ik}\lbrace \Pi_k,g\rbrace.
\label{vv-6}
\end{eqnarray} 
The result of the Dirac procedure is the elimination of the momenta 
conjugate to the fermionic variables leaving us with the following non-zero 
Dirac bracket relations
\begin{eqnarray}
\lbrace R, \pi_R \rbrace^\ast &=& \lbrace R, \pi_R \rbrace = - 1,
\qquad \lbrace z_A, \pi_z^B \rbrace^\ast = 
\lbrace z_A, \pi_z^B \rbrace = - \delta_A^B,\nonumber\\
\lbrace \bar z_{\bar A}, \pi_{\bar z}^{\bar B} \rbrace^\ast &=& 
\lbrace \bar z_{\bar A}, \pi_{\bar z}^{\bar B} \rbrace =  
- \delta_{\bar A}^{\bar B}, \qquad 
\lbrace \chi^A, \bar\chi^{\bar B} \rbrace^\ast = -i\kappa^2 G^{A \bar B},
\label{vv-7} \\
\lbrace \phi^A, \bar\phi^{\bar B} \rbrace^\ast &=& -i\kappa^2 G^{A \bar B}, 
\qquad\qquad \lbrace \lambda, \bar\lambda \rbrace^\ast = 
\frac{3}{2}i.
\nonumber
\end{eqnarray}
The canonical Hamiltonian is the sum of all the first-class constraints
\begin{eqnarray}
H_c = NH + i\frac{\bar\psi}{2} S + i\frac{\psi}{2} \bar S + 
\frac{V}{2}{\cal F}, \label{vv-8}
\end{eqnarray}
where $H$ is the classical Hamiltonian of the system written as
\begin{eqnarray}
H &=& -\frac{\kappa^2}{12 R}\pi_R^2 + R^3 U(R, z, \bar z) - \frac{\sqrt k}
{3R}\bar\lambda \lambda + \frac{\kappa^2}{R^3} \bar p_{\bar A}
G^{\bar A B} p_B - \frac{i\kappa}{2R^3}(\bar\lambda \bar\chi^{\bar A} + 
\lambda \bar\phi^{\bar A}) \bar p_{\bar A} \nonumber \\
&-& \frac{i\kappa}{2R^3}(\bar\lambda \phi^A + \lambda\chi^A)p_A + 
\frac{1}{\kappa^2 R^3} R_{\bar A B \bar C D} \bar\chi^{\bar A} \chi^B
\bar\phi^{\bar C}\phi^D + \frac{1}{\kappa} e^{\frac{\alpha G}{2}}
\bar\lambda \lambda \nonumber \\
&+& \frac{1}{4R^3}\bar\lambda \lambda G_{\bar A B}
(\bar\chi^{\bar A}\chi^B + \phi^B \bar\phi^{\bar A}) 
- \frac{3}{16\kappa^2 R^3} \lbrack G_{\bar A B}(\bar\chi^{\bar A B} \chi^B
+ \phi^B \bar\phi^{\bar A})\rbrack^2 
\label{vv-9} \\
&-& \frac {3\sqrt k}{2\kappa^2 R} G_{\bar A B}
(\bar \chi^{\bar A} \chi^B + \phi^B \bar \phi^{\bar A}) +
\frac{3}{2\kappa^3} e^{\frac{\alpha G}{2}}G_{\bar A B}
(\bar \chi^{\bar A} \chi^B + \phi^B \bar\phi^{\bar A}) + \frac{2}{\kappa^3}
(e^{\frac{\alpha G}{2}})_{, AB} \chi^A \phi^B \nonumber \\
&+& \frac{2}{\kappa^3}(e^{\frac{\alpha G}{2}})_{, \bar A \bar B} 
\bar\phi^{\bar B} \bar\chi^{\bar A} +
\frac{2}{\kappa^3}(e^{\frac{\alpha G}{2}})_{,\bar A B}(\bar\chi^{\bar A}
\chi^B + \phi^B \bar\phi^{\bar A}) + \frac{1}{\kappa^2} \bar\lambda
\lbrack (e^{\frac{\alpha G}{2}})_{, A} \phi^A + 
(e^{\frac{\alpha G}{2}})_{,\bar A} \bar\chi^{\bar A} \rbrack \nonumber\\
&-& \frac{1}{\kappa^2}\lambda
\lbrack (e^{\frac{\alpha G}{2}})_{, A} \chi^A + 
(e^{\frac{\alpha G}{2}})_{,\bar A}
\bar\phi^{\bar A} \rbrack ,\nonumber   
\end{eqnarray}
$S$ and $\bar S$ are the classical supersymmetric generators of our model
\begin{eqnarray}
S &=& \left[\frac { \kappa}{3 \sqrt R} \pi_R + \frac{2i}{\kappa} \sqrt k 
\sqrt R - \frac{2i}{\kappa^2} \sqrt{R^3} e^\frac{\alpha G}{2} + 
\frac{i}{2 \kappa\sqrt{R^3}} G_{\bar A B}(\bar\chi^{\bar A} \chi^B + 
\phi^B \bar\phi^{\bar A})
\right]\lambda \label{vvv} \\ 
&+& \left[\frac {1}{\sqrt{R^3}}p_C - \frac{2i}{\kappa^3} \sqrt{R^3}
(e^\frac {\alpha G}{2})_{, C} \right] \phi^C + 
\left[ \frac {1}{\sqrt {R^3}} \bar p_{\bar C} - \frac {2i}{\kappa^3}
\sqrt {R^3} (e^\frac{\alpha G}{2})_{,\bar C} \right]\bar\chi^{\bar C},
 \nonumber 
\end{eqnarray}
\begin{eqnarray}
\bar S &=& \left[\frac {\kappa}{3\sqrt R} \pi_R - \frac{2i}{\kappa} 
\sqrt k \sqrt R + \frac{2i}{\kappa^2} \sqrt {R^3} e^\frac{\alpha G}{2} - 
\frac{i}{2\kappa \sqrt {R^3}} G_{\bar A B}(\bar\chi^{\bar A} \chi^B +
\phi^B \bar\phi^{\bar A}) \right] \bar \lambda \label{vvv-1} \\
&+& \left[\frac{1}{\sqrt {R^3}} \bar p_{\bar C} + \frac{2i}{\kappa^3}
\sqrt {R^3}(e^\frac{\alpha G}{2})_{,\bar C} \right] \bar\phi^{\bar C} +
\left[\frac{1}{\sqrt {R^3}} p_C + \frac{2i}{\kappa^3} \sqrt {R^3}
(e^\frac{\alpha G}{2})_{, C} \right] \chi^C ,\nonumber
\end{eqnarray}  
and ${\cal F}$ is the classical $U(1)$ rotation generator 
\begin{eqnarray}
{\cal F} &=& -\frac{2}{3} \bar\lambda \lambda + \frac {G_{\bar A B}}
{\kappa^2}(\bar\chi^{\bar A} \chi^B + \phi^B \bar\phi^{\bar A}).
\label{vvv-2}
\end{eqnarray}
These first-class constraints are obtained from the component action 
(\ref{ocho}) varying $N(t)$, $\psi(t)$, $\bar\psi(t)$ and $V(t)$. 

We see from (\ref{vv-1}), that the canonical momenta are written as

\begin{eqnarray}
p_A(\bar z) &=& \pi_A + \frac{i}{2\kappa^2} G_{\bar M B}(\Gamma_{A D} ^B 
\bar\chi^{\bar M}\chi^D + \Gamma_{A D}^B \bar\phi^{\bar M}\phi^D), 
\label{vvv-3} \\ 
\bar p_{\bar A}(z) &=& \bar \pi_{\bar A} - \frac{i}{2\kappa^2} G_{\bar M B}
(\Gamma_{\bar A \bar D} ^{\bar M} \bar \chi^{\bar D}\chi^B + 
\Gamma_{\bar A \bar D} ^{\bar M} \bar \phi^{\bar D} \phi^B), \nonumber 
\end{eqnarray}
for which we have the following non-zero Dirac brackets

\begin{eqnarray}
\lbrace p_A, \bar p_{\bar B} \rbrace^\ast &=& -\frac{i}{\kappa^2}
R_{A \bar B C \bar D}(\chi^C \bar\chi^{\bar D} + \phi^C \bar\phi^{\bar D}),
\nonumber \\
\lbrace p_A, \bar \chi_{\bar C} \rbrace^\ast &=& \frac{1}{2}G_{\bar C B}
\Gamma^B_{AD}\chi^D, \qquad\qquad
\lbrace \bar p_{\bar A}, \chi_C \rbrace^\ast = \frac{1}{2}G_{A B}
\Gamma^{\bar B}_{\bar A \bar D}\bar\chi^{\bar D}, \label{vvv-4} \\
\lbrace p_{A}, \bar\phi_{\bar C} \rbrace^\ast &=& \frac{1}{2}G_{\bar C B}
\Gamma^{B}_{AD}\phi^{D}, \qquad\qquad
\lbrace \bar p_{\bar A}, \phi_C \rbrace^\ast = \frac{1}{2}G_{C \bar B}
\Gamma^{\bar B}_{\bar A \bar D}\bar \phi^{\bar D}. \nonumber 
\end{eqnarray}

In terms of those Dirac brackets (\ref{vv-7}) and (\ref{vvv-4}) we have a 
closed super-algebra of the conserving even charges $H, {\cal F}$ and odd 
supercharges
$S, \bar S$, 
\begin{eqnarray}
\lbrace S, \bar S \rbrace^\ast &=& - 2iH, \qquad 
\lbrace S, H \rbrace^\ast = \lbrace \bar S, H \rbrace^\ast = 0,
\qquad \lbrace {\cal F}, S \rbrace^\ast = iS, \label{vvv-5} \\ 
\lbrace {\cal F}, \bar S \rbrace^\ast &=& -i\bar S, \qquad 
\lbrace {\cal F}, H \rbrace^\ast = 0, \qquad 
\lbrace S, S \rbrace^\ast = 0,\qquad
\lbrace \bar S, \bar S \rbrace^\ast = 0. \nonumber
\end{eqnarray}

\section {Quantization of the model}

On the quantum level we replace the Dirac brackets (\ref{vv-7}) by 
anticommutators if both arguments are odd, and by commutators if
they are both even
\begin{equation}
\lbrace O_1, O_2 \rbrace = i\lbrace O_1, O_2 \rbrace^{\ast}, \qquad
\lbrack E, O \rbrack = i\lbrack E, O\rbrack^{\ast},\qquad
\lbrack E_1, E_2 \rbrack = i\lbrack E_1, E_2 \rbrack^{\ast},
\label{vvv-6}
\end{equation}
in particular, this gives the following non-zero commutation relations 
for the Dirac brackets 
\begin{eqnarray}
\lbrack R, \pi_R \rbrack &=& -i, \qquad \lbrack z^A, \pi_B \rbrack = 
-i\delta^A_B, \qquad \lbrack \bar z^{\bar A}, \bar\pi_{\bar B} \rbrack = 
-i\delta^{\bar A}_{\bar B}, \label{vvv-7}\\
\lbrace \lambda, \bar\lambda \rbrace &=& -\frac{3}{2}, \qquad  
\lbrace \chi^A, \bar\chi_B \rbrace = \kappa^2 \delta^A_B,\qquad  
\lbrace \phi^A, \bar\phi_B \rbrace = \kappa^2 \delta^A_B, 
\nonumber
\end{eqnarray}
where $\bar\lambda, \bar\chi_{\bar B}$ and $\bar\phi_{\bar B}$ are 
hermitian conjugates to $\lambda, \chi_B$ and $\phi_B$ with respect to 
operation $(\chi^A)^{\dagger}= \bar\chi_A, (\phi^B)^{\dagger}= 
\bar\phi_B$, $\bar\chi_A G^{A \bar A}=\bar\chi^{\bar A}$ and
$\bar\phi_A G^{A \bar A} = \bar\phi^{\bar A}$. In the quantum theory the 
first class-constraints (\ref{vv-9}-\ref{vvv-2}) 
associated with the invariance of the action (\ref{v-9}) become conditions 
on the wave function $\Psi(R, z, \lambda, \chi, \phi)$. So, that any 
physically allowed states must obey the following quantum constraints

\begin{equation}
H|\Psi>= 0,\qquad S|\Psi>=0, \qquad \bar S|\Psi>=0, \qquad {\cal F}|\Psi>=0, 
\label{vvv-8}
\end{equation}
where the first equation in (\ref{vvv-8}) is the so-called Wheeler-DeWitt 
equation for minisuperspace models.

The quantum generators $H, S, \bar S$ and ${\cal F}$ form a closed 
super-algebra of the supersymmetric quantum mechanics under the Dirac 
brackets (\ref{vv-7}) and (\ref{vvv-4}). 
\begin{eqnarray}
\lbrace S, \bar S \rbrace &=& 2H, \qquad \lbrack S, H \rbrack = 
\lbrack \bar S, H \rbrack = 0, \qquad \lbrack {\cal F}, S \rbrack = -S, 
\label{vvv-9} \\
\lbrack {\cal F}, \bar S \rbrack &=& -\bar S,\qquad
\lbrack {\cal F}, H \rbrack = 0,\qquad\qquad\qquad S^2 = \bar S^2 = 0, 
\nonumber
\end{eqnarray}
where $H$ is the Hamiltonian, $S$ is the single complex supersymmetric 
operator and ${\cal F}$ is the fermion number operator. 
The super-algebra
(\ref{vvv-9}) doesn't define a positive-definite Hamiltonian.

In the usual canonical quantization the even canonical variables are 
replaced by operators 
\begin{equation}
R \to R, \qquad \pi_R = i\frac{\partial}{\partial R},\qquad
z^A \to z^A, \qquad \pi_A = i\frac{\partial}{\partial z^A}, 
\label{vvv-10}
\end{equation}
and the odd variables $\lambda, \bar\lambda, \chi^A, \bar\chi^A,
\phi^A$ and $\bar\phi^A$, which obey the Dirac brackets (\ref{vv-7})
after quantization become anticommutators. We can fulfill them on the Fock 
space representation with $\bar\lambda, \bar\chi_A, \bar\phi_A$
as a creation, and $\lambda, \chi^A$ and $\phi^A$ as annihilation operators
with $|0>$ vacuum of the Fock space, so that 
$\lambda|0>= \chi_A|0>=\phi_A|0>=0$ and the general quantum states can be
written as the vectors depending on $R, z_A,$ and $\bar z_{\bar A}$ in the
corresponding Fock space. There is other approach \cite{diesiocho}, which 
ensures the canonical anticommutation rules (\ref{vvv-7}). Now, we can write 
$\lambda, \bar\lambda, \chi_A, \bar\chi_A, \phi_A$ and 
$\bar\phi_A$ in the form of the direct product of $1 + 2n$ matrix
$2 \times 2$, then, we obtain the following matrix realization for the case 
of $n$ complex matter supermultiplets 

\begin{eqnarray}
\lambda &=& \sqrt{\frac{3}{2}} \sigma_1^{(-)}\otimes1_2\otimes.....
\otimes1_{2n+1}, \qquad
\lambda^{\dagger} = \sqrt{\frac{3}{2}} \sigma_1^{(+)}\otimes1_2\otimes.....
\otimes1_{2n+1}, \nonumber\\
\phi^i &=& \kappa \sigma_1^{(3)}\otimes.....\otimes\sigma_{2i-1}^{(3)}\otimes
\sigma_{2i}^{(-)}\otimes1_{2i+1}\otimes.....\otimes1_{2n+1}, \nonumber\\
\bar\phi_i &=& \kappa \sigma_1^{(3)}\otimes.....\otimes\sigma_{2i-1}^{(3)}
\otimes \sigma_{2i}^{(+)}\otimes1_{2i+1}\otimes.....\otimes1_{2n+1},
\label{vvv-11} \\
\chi^i &=& \kappa \sigma_1^{(3)}\otimes.....\otimes\sigma_{2i}^{(3)}
\otimes \sigma_{2i+1}^{(-)}\otimes1_{2i+2}\otimes.....\otimes1_{2n+1}, 
\nonumber\\
\bar\chi_i &=& \kappa \sigma_1^{(3)}\otimes.....\otimes\sigma_{2i}^{(3)}
\otimes \sigma_{2i+1}^{(+)}\otimes1_{2i+2}\otimes.....\otimes1_{2n+1},
\nonumber
\end{eqnarray}
where the down index in the direct product of the matrix shows place of the 
matrix $(\it i=1,2.....,n)$, $\sigma^{\pm} = 
\frac{\sigma^1 \pm \sigma^2}{2}$ 
with $\sigma^1, \sigma^2$ and $\sigma^3$ Pauli matrices.

When classical variables $H, S, S^{\dagger}$ and $\cal F$ become operators 
on the quantum level we must consider the nature of the Grassmann variables
$\lambda, \lambda^{\dagger}, \chi^A, \bar\chi_A, \phi^A$ and
$\bar\phi_A$, and with respect to those ones we perform the 
antisymmetrization, $i.e$, we can write the bilinear combination in the 
form of commutators, $e.g$ $\lambda^{\dagger}\lambda = 
\frac{1}{2}\lbrack \lambda^{\dagger}, \lambda \rbrack$. To obtain the 
quantum expression for the hamiltonian $H$ and for the supercharges $S$ and 
$S^{\dagger}$ we must solve the operator ordering ambiguity. Such 
ambiguities always arise, when the operator expression contains the 
product of non-commutating operators $R, \pi_R, z^A$ and $\pi_A$ in our 
case \cite{diesiocho}. 
Technically it means the following: for the quantum supercharges we take 
the same order that for the operator in (\ref{vv-9}). Then, we must 
integrate with measure 
$R^{1/2}(\det G_{A \bar B})^{1/2} dR d^n z d^n \bar z $ in the inner
product of two states. In this measure the momenta hermitian-conjugate
$\pi_R = i\frac{\partial}{\partial R}$ is non-hermitian with 
$\pi_R^{\dagger} = R^{-1/2} \pi_R R^{1/2}$, however, the combination 
$(R^{-1/2}\pi_R)^{\dagger} = \pi_R^{\dagger}R^{-1/2} = R^{-1/2} \pi_R$ is
hermitian. The canonical momenta $\pi_A^{\dagger}$ hermitian-conjugate to
$\pi_A = i\frac{\partial}{\partial z^A}$ have the form
$(\pi_A)^{\dagger} = g^{-1/2}(\bar\pi_{\bar A})g^{1/2}$,    
where $g = \det G_{A \bar B}$ and $\bar\pi_{\bar A} = 
i\frac{\partial}{\partial \bar z^{\bar A}}$,
such a procedure leads in our case to the following expression for 
supercharges $S, S^{\dagger}$ and the fermionic number operator ${\cal F}$

\begin{eqnarray}
S &=& \lbrace \frac{\kappa}{3}R^{-1/2} \pi_R + \frac{2i}{\kappa}
\sqrt k R^{1/2} - \frac{2i}{\kappa^2}R^{3/2}e^{\frac{\alpha G}{2}} + 
\frac{iR^{-3/2}}{4\kappa} \lbrack \bar\chi_A, \chi^A \rbrack + 
\frac{iR^{-3/2}}{4\kappa} \lbrack \phi^A, \bar\phi_A \rbrack 
\rbrace \lambda \label{vvv-12}\\ 
&+& \lbrack R^{-3/2} p_C - \frac{2iR^{3/2}}{\kappa^3} 
(e^{\frac{\alpha G}{2}})_{,C} \rbrack \phi^C +
\lbrack R^{-3/2} \bar p^C - \frac{2iR^{3/2}}{\kappa^3} 
(e^{\frac{\alpha G}{2}}),^C \rbrack \bar\chi_C, \nonumber
\end{eqnarray}

\begin{eqnarray}
S^{\dagger} &=& \lbrace \frac{\kappa}{3}R^{-1/2} \pi_R - 
\frac{2i}{\kappa}\sqrt k R^{1/2}
+ \frac{2i}{\kappa^2}R^{3/2}e^{\frac{\alpha G}{2}} - \frac{iR^{-3/2}}
{4\kappa} \lbrack \bar\chi_A, \chi^A \rbrack - 
\frac{iR^{-3/2}}{4\kappa} \lbrack \phi^A, \bar\phi_A \rbrack \rbrace 
\lambda^{\dagger} \label{vvv-13}\\
&+& \lbrack R^{-3/2} \bar p^C + \frac{2iR^{3/2}}{\kappa^3} 
(e^{\frac{\alpha G}{2}}),^C \rbrack \bar\phi_C +
\lbrack R^{-3/2} p_C + \frac{2iR^{3/2}}{\kappa^3} 
(e^{\frac{\alpha G}{2}})_{, C} \rbrack \chi^C, \nonumber
\end{eqnarray}

\begin{equation}
{\cal F} = \frac{1}{2}(\frac{2}{3} \lbrack \lambda^{\dagger}, \lambda 
\rbrack + \frac{1}{\kappa^2} \lbrack \bar\chi_A, \chi^A \rbrack +
\frac{1}{\kappa^2} \lbrack \phi^A, \bar\phi_A \rbrack),
\label{vvv-14}
\end{equation}
where $p_A = i\frac{\partial}{\partial z^A} + \frac{i}{4\kappa^2}
\Gamma_{A D}^B (\lbrack \bar\chi_B, \chi^D \rbrack + 
\lbrack \bar\phi_B, \phi^D \rbrack)$, 
$\bar p^C = \bar p_{\bar B}G^{\bar B C}$ and 
$( )_,^C = ( )_{,\bar B}G^{\bar B C}$. Note, that $(p_A)^{\dagger} = 
g^{-1/2}\bar p_{\bar A}g^{1/2}$, then, the anticommutation
relation $\lbrace S, S^{\dagger}\rbrace = 2H$ and 
$S^2 = S^{\dagger 2}=0$ fix all additional terms and define the quantum 
Hamiltonian, but in this case the operational expression 
$\frac{\kappa^2}{12}(R^{-1/2} \pi_R R^{-1/2}\pi_R)$
corresponding to the energy of the scale factor $R$ contributes to 
positive in the Hamiltonian, as well as to the energy of the scalar fields. 
As we can see from classical Hamiltonian (\ref{vv-9}) the energy of the 
scale factor is negative, this is due to the fact, that the particle-like 
fluctuations don't correpond to the scale factor. This is reflected in 
the fact, that the anticommutator value 
$\lbrace \lambda, \bar\lambda \rbrace = -\frac{3}{2}$ 
of superpartners $\lambda, \bar\lambda$ of the scale factor $R$ is 
negative, unlike anticommutation relations (\ref{vv-7}), which are 
positive. Anticommutation relations may be fulfilled under the conditions

\begin{equation} 
\bar \lambda = -\lambda^{\dagger} \qquad (\chi^A)^{\dagger} = 
\bar\chi_A \qquad (\phi^A)^{\dagger} = \bar\phi_A,
\label{vvv-15}
\end{equation}
where $\lbrace \lambda, \lambda^{\dagger} \rbrace = \frac{3}{2}$. Then, the 
equation may be written in the form
\begin{equation}
\bar\lambda = \xi^{-1}\lambda^{\dagger}\xi,\qquad \bar\chi_A = 
\xi^{-1}(\chi^A)^{\dagger}\xi, \qquad \bar\phi_A=
\xi^{-1}(\phi^A)^{\dagger}\xi.
\label{vvv-16}
\end{equation}
In order to have a consisntence with the expression (\ref{vvv-15}) and 
(\ref{vvv-16}) it is necessary, that the operator $\xi$ possess the 
following properties 
\begin{equation}
\lambda^{\dagger}\xi = -\xi\lambda^{\dagger}, \qquad 
(\chi^A)^{\dagger}\xi = \xi (\chi^A)^{\dagger},\qquad (\phi^A)^{\dagger} 
\xi = \xi(\phi^A)^{\dagger}. \label{vvv-17}
\end{equation}
The operators $\bar\lambda , \bar\chi_A$ and $\bar\phi_A$ will be
conjugate to operators $\lambda, \chi^A$ and $\phi^A$ under the inner 
product of two states
\begin{equation}
<\Psi_1, \Psi_2>_{\xi} = \int \Psi_1^\ast \xi \Psi_2 R^{1/2}g^{1/2}
dRd^n z d^n \bar z,
\label{vvv-18} 
\end{equation}
which in general is non-positive. In the matrix realization the 
operator $\xi$ has the form
\begin{equation}
\xi= \sigma_1^{(3)}\otimes 1_2\otimes.....\otimes1_{2n+1},
\label{vvv-19}
\end{equation}
and it can be written as a difference of two projection operators
$p_+ = \frac{1}{2}(1+\xi)$ and $p_{-} = \frac{1}{2}(1- \xi)$. On the other
hand, when the states fulfill equation (\ref{vvv-8}) with zero energy on 
subspace of Fock space with vacuum $p_+|0> = |0>$, the 
inner product (\ref{vvv-18}) is positive-definite.

So, for the supercharge operator $S$ we can construct 
conjugation (\ref{vvv-18}) under the operator $\bar S$ with the help 
of the following equation
\begin{equation}
\bar S = \xi^{-1} S^{\dagger} \xi.
\label{4v-1}
\end{equation}
In the general case any arbitrary operator ${\cal L}$ conjugate with 
respect to (\ref{vvv-18}) has the form $\bar {\cal L} = 
\xi^{-1}{\cal L}^{\dagger} \xi$, the same operator $\bar \xi$ is conjugate 
$\bar\xi = \xi^{-1} \xi^{+}\xi$, if the condition $\xi = \xi^{\dagger}$ is 
fulfilled.  

We can see, that the anticommutator of supercharge $S$ and their conjugate
$\bar S$ under our conjugate operation has the form
\begin{equation}
\overline{\lbrace S, \bar S\rbrace} = 
\xi^{-1} \lbrace S, \bar S \rbrace^{\dagger} \xi = \lbrace S, \bar S \rbrace,
\label{4v-2}
\end{equation} 
and it is a self-conjugate operator.

As a consecuence of the algebra (\ref{vvv-9}) we obtain, that the 
Hamiltonian $H$ is a self-conjugate operator 
$\bar H = \xi^{-1} H^{\dagger} \xi = H$ and its value will be real. 
Then, the quantum Hamiltonian will have the form of the classical 
Hamiltonian (\ref{vv-9}) with antisymmetrization under 
fermionic operator and with representation (\ref{vvv-4}) for the momenta 
covariant operator $p_A$ and $\bar p_{\bar A}$. The kinetical term for the 
scale factor $R$ and matter fields will have the following order form

\begin{equation}
-\frac{\kappa^2}{12}R^{-1/2}\pi_R R^{-1/2} \pi_R + \frac{\kappa^2}{R^3}
g^{-1/2} \bar p_{\bar A}g^{1/2} G^{\bar A B} p_B.
\label{4v-3}
\end{equation}      
Note, that the super-algebra (\ref{vvv-9}) does not define 
positive-definite Hamiltonian in a full agreement with the circumstance,
that the potential $V(z, \bar z)$ of scalar fields is not 
positive semi-definite in contrast with the standard quantum mechanics. 

\section{Conclusion}
The Grassmann components of the vacuum configuration with the FRW
metric may be obtained by decomposition of the Rarita-Schwinger field in the
following way: commuting covariant constant spinors $\lambda_{\alpha}(x^i)$
and $\bar\lambda_{\dot \alpha}(x^i)$ are fixed on the configuration space,  
and on the other hand, time-like depending variables are not spinors. The 
time-like components of the Rarita-Schwinger field may be written as
\begin{equation}
\psi_0^{\alpha}(x^i,t) = \lambda^{\alpha}(x^i) \psi(t), \qquad
\bar\psi_0^{\dot\alpha}(x^i,t) = \bar\lambda^{\dot\alpha}(x^i) \bar\psi(t).
\end{equation}

The spatially components of the Rarita-Schwinger field has the following
representation corresponding to the direct product time-subspace on three-
space of the fixed spatial configuration (in our case it is a plane or
a three sphere). Expicitly we get
\begin{equation}
\psi_m^{\alpha}(x^i,t) = e_m^{(\mu)}\sigma_{(\mu)}^{\alpha \dot\beta}
\bar\lambda_{\dot\beta}(x^i) \bar\lambda(t), \qquad
\bar\psi_m^{\dot\alpha}(x^i,t) = 
e_m^{(\mu)}\sigma_{(\mu)}^{\dot\alpha \beta}\lambda_{\beta}(x^i)\lambda(t),
\end{equation} 
where $e_m^{(\mu)}(x^i,t)$ are the tetrad for the FRW metric. Those fermion
representation are solutions of the supergravity equations.

Hence, specific quantum supersymmetric mechanics corresponding to quantum 
level in our models define the structure, which permits the fundamental 
quantum states invariant under the $n=2$ local conformal supersymmetry 
in $N=1$ supergravity interacting with a set of matter fields 
\cite{diesisiete}. In our case the small supersymmetry is a subgroup of 
the space-time supersymmetry. In the small supersymmetry the parameter 
$\alpha$ is not necessarily $\alpha =1$, and the mechanism of spontaneous 
breaking of local small supersymmetry induces the general mechanism of 
spontaneous breaking of supersymmetry in the supergravity theories.

In our case the constraints on the wave function of the universe permit 
the existence of non-trivial solutions, unlike the standard formulation on 
minisuperspace models, in which the Lorentz constraint is present 
\cite{diez}. The Lorentz constraints imposes many conditions on the wave 
function, which can lead to trivial solutions \cite{diez,diesinueve}.

The next step is to find non-trivial wave function of the universe for 
different set of fields including dilaton in the spontaneous breaking 
phase, as well as to establish dependence of the universe parameters with 
parameter of spontaneous breaking of supersymmetry gravitino mass. This 
wave function will be a vector-state with zero energy in the supergravity 
theories or in the effective superstring theory. 

\vspace {.5 cm}
{\bf Acknowledments}: We are grateful to A. Cabo, E. Ivanov,
S. Krivonos, I. Lyanzuridi, L. Marsheva, O. Obreg\'on, A. Pashnev and 
M.P. Ryan  for their interest in this paper. This work was supported in 
part by CONACyT grant 3898P-E9608.

\end {document}